\documentclass[prb,preprint,aps,superscriptaddress,longbibliography]{revtex4-1}

\usepackage{graphicx, epstopdf}
\usepackage{subfigure}
\usepackage{float}
\usepackage{amsmath}
\usepackage{amsfonts}
\usepackage{amssymb}
\usepackage{bm}
\usepackage{subfigure}
\usepackage{caption}
\usepackage{hyperref}
\usepackage{color}
\begin{document}

\title{Effects of intervalley scatterings in thermoelectric performance of band-convergent antimonene}

\author{Yu Wu}
\affiliation{Key Laboratory of Micro and Nano Photonic Structures (MOE) and Key Laboratory for Information Science of Electromagnetic Waves (MOE) and Department of Optical Science and Engineering, Fudan University, Shanghai 200433, China}
\author{Bowen Hou}
\affiliation{Key Laboratory of Micro and Nano Photonic Structures (MOE) and Key Laboratory for Information Science of Electromagnetic Waves (MOE) and Department of Optical Science and Engineering, Fudan University, Shanghai 200433, China}
\author{Congcong Ma}
\affiliation{Department of Light Sources and Illuminating Engineering, and Academy for Engineering \& Technology, Fudan University, Shanghai, 200433, China}
\author{Jiang Cao}
\affiliation{School of Electronic and Optical Engineering, Nanjing University of Science and Technology, Nanjing 210094, China}
\author{Ying Chen}
\affiliation{Department of Light Sources and Illuminating Engineering, and Academy for Engineering \& Technology, Fudan University, Shanghai, 200433, China}
\author{Zixuan Lu}
\affiliation{Department of Light Sources and Illuminating Engineering, and Academy for Engineering \& Technology, Fudan University, Shanghai, 200433, China}
\author{Haodong Mei}
\affiliation{Key Laboratory of Micro and Nano Photonic Structures (MOE) and Key Laboratory for Information Science of Electromagnetic Waves (MOE) and Department of Optical Science and Engineering, Fudan University, Shanghai 200433, China}
\author{Hezhu Shao}
\affiliation{College of Electrical and Electronic Engineering, Wenzhou University, Wenzhou, 325035, China}
\author{Yuanfeng Xu}
\email{xuyuanfeng19@sdjzu.edu.cn}
\affiliation{School of Science, Shandong Jianzhu University, Jinan 250101, Shandong, China}
\author{Heyuan Zhu}
\affiliation{Key Laboratory of Micro and Nano Photonic Structures (MOE) and Key Laboratory for Information Science of Electromagnetic Waves (MOE) and Department of Optical Science and Engineering, Fudan University, Shanghai 200433, China}
\author{Zhilai Fang}
\affiliation{Department of Light Sources and Illuminating Engineering, and Academy for Engineering \& Technology, Fudan University, Shanghai, 200433, China}
\author{Rongjun Zhang}
\email{rjzhang@fudan.edu.cn}
\affiliation{Key Laboratory of Micro and Nano Photonic Structures (MOE) and Key Laboratory for Information Science of Electromagnetic Waves (MOE) and Department of Optical Science and Engineering, Fudan University, Shanghai 200433, China}
\author{Hao Zhang}
\email{zhangh@fudan.edu.cn}
\affiliation{Key Laboratory of Micro and Nano Photonic Structures (MOE) and Key Laboratory for Information Science of Electromagnetic Waves (MOE) and Department of Optical Science and Engineering, Fudan University, Shanghai 200433, China}
\affiliation{Nanjing University, National Laboratory of Solid State Microstructure, Nanjing 210093, China}

\date{\today}

\begin{abstract}
The strategy of band convergence of multi-valley conduction bands or multi-peak valence bands has been widely used to search or improve thermoelectric materials. However, the phonon-assisted intervalley scatterings due to multiple band degeneracy are usually neglected in the thermoelectric community. In this work, we investigate the (thermo)electric properties of non-polar monolayer $\beta$- and $\alpha$-antimonene considering full mode- and momentum-resolved electron-phonon interactions. We also analyze thoroughly the selection rules on electron-phonon matrix-elements using group-theory arguments. Our calculations reveal strong intervalley scattering between the nearly degenerate valley states in both $\beta$- and $\alpha$-antimonene, and the commonly-used deformation potential approximation neglecting the dominant intervalley scattering gives inaccurate estimations of the electron-phonon scattering and thermoelectric transport properties. By considering full electron-phonon interactions based on the rigid-band approximation, we find that, the maximum value of the thermoelectric figure of merits $zT$ at room temperature reduces to 0.37 in $\beta$-antimonene, by a factor of 5.7 comparing to the value predicted based on the constant relaxation-time approximation method. Our work not only provides an accurate prediction of the thermoelectric performances of antimonenes that reveals the key role of intervalley scatterings in determining the electronic part of zT, but also showcases a computational framework for thermoelectric materials.
\end{abstract}

\flushbottom
\maketitle

\thispagestyle{empty}


The thermoelectric (TE) performance of a material is quantified by a dimensionless figure of merit $zT$ ($zT=S^2\sigma T/(\kappa_{e}+\kappa_l)$), where $S$ represents the Seebeck coefficient, $\sigma$ is the electron conductivity, and $\kappa_{e/l}$ is the electronic/lattice thermal conductivity. Good TE materials generally possess good power factor $PF$ $(PF=S^2\sigma)$ and poor thermal conductivity $\kappa$ $(\kappa=\kappa_{e}+\kappa_l)$. However, the three electronic properties $S$, $\sigma$ and $\kappa_e$ are tightly coupled together, e.g. an increasing carrier concentration leads to increasing $\sigma$ and decreasing $S$ simultaneously, thus making optimization of the TE performance a daunted task. As a relatively independent quantity in $zT$, much effort has been put to reduce the $\kappa_l$ and thus to enhance $zT$\cite{Kaur2019,Hanus2019,WuYixuan2019}. Past decades have witnessed the significant improvements in $zT$s, owning to the discovery of many new high-performance TE materials and several newly-emergent strategies to improve $zT$s.  On the other hand, it is noted the Seebeck coefficient $S$ is proportional to the density-of-states effective mass $m^*=N_v^{2/3}m_b^*$ where $N_v$ is the band degeneracy and $m_b^*$ is the band effective mass\cite{Pei2011,Zeier2016}, while $\sigma$ is inversely related with the band effective mass $m^*_b$. Hence, high $N_v$ and low $m_b^*$ is beneficial for TE performance by simultaneously achieving high $\sigma$ and $S$. 

The $\beta$-antimonene (Sb) possesses complex layer-number-dependent electronic properties, which undergoes a topological transformation from a topological semimetal to a topological insulator at 22 bilayers, then to a quantum spin Hall (QSH) phase at 8 bilayers, and finally to a topologically trivial semiconductor with narrow bandgap when thinning down to 3 bilayers or thinner\cite{Zhang2012}. Zhou \textit{et al.} proposed an efficient way to realize both the large intrinsic QSH and anomalous Hall conductivity in a single honeycomb Sb monolayer grown on a ferromagnetic MnO$_2 $(H-MnO$_2 $) layer\cite{Zhou2016,Zhang2018}. Experimentally, monolayer and few-layer $\beta$-Sb have been successfully synthesized                                                                                                                                                                                                                                                                                                                                                                                                                                                                                                                                                                                                                                                                                                                                                                                                                                                                                                                                                                                                                                                                                                                                                                                                                                                                                                                                                                                                                                                                                                                                                                                                                                                                                                                                                                                                                           with large scale and high quality, which can be applied in thermophoto-voltaic devices, perovskite solar cell as a hole transport layer material and electrocatalysis technologies\cite{Wang2018}. Compared with $\beta$ structure, it is still challenging to fabricate $\alpha$-Sb although small patches of $\alpha$-Sb have been observed on islands of $\alpha$-Bi\cite{Maerkl2017}. Recently, Shi \textit{et al} successfully synthesized  $\alpha$-Sb on the substrate of bulk T$_d$-WTe$_2$ with tunable layers by using the molecular beam epitaxy (MBE) technique\cite{zhiqiang2019}. Monolayer $\alpha$-Sb is chemically stable when exposed to ambient environment and possesses good electrical conductivities.

\begin{figure}
\centering
\includegraphics[width=1\linewidth]{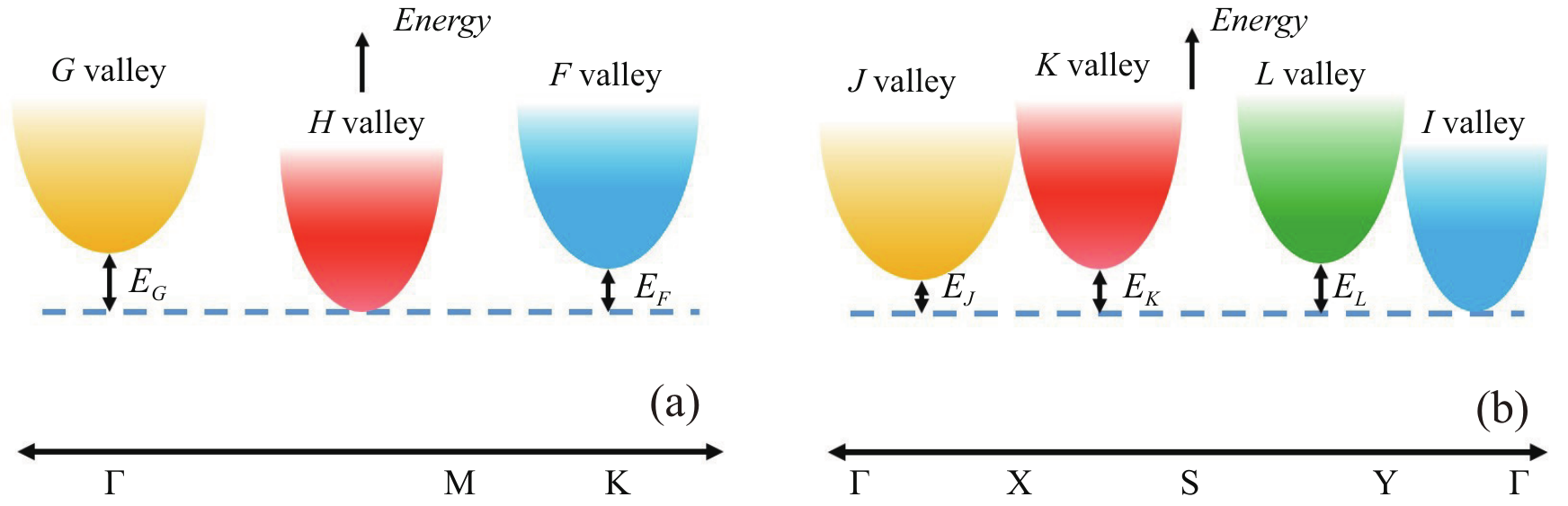}
\caption{Schematic of multiple valley pockets in the conduction bands for (a) $\beta$-Sb and (b) $\alpha$-Sb respectively.}
\label{sketch}
\end{figure}

The crystal and electronic band structure of monolayer Sb are shown in Fig.~S1 and Fig.~S2(a, b). Both monolayer $\alpha$- and $\beta$-Sb are non-polar and possess convergence of multi-valley conduction bands attributed to the orbital degeneracy and crystal symmetry, which is believed to be beneficial to the high TE performance in materials\cite{Pei2011,wei2012,Pei2012a,Tang2015,Zeier2016,Fang2019}. As shown in Fig.~\ref{sketch}, the conduction bands of $\beta$-Sb possess the local band-minima of $G$ and $F$ besides the global CBM of $H$, with respective minimum energies of $E_{F}=250\;\rm{meV}$ and $E_{G}=290\;\rm{meV}$ above the CBM of $V_H$ along the $\Gamma-\rm{M}$ direction. For monolayer $\alpha$-Sb, more energy valleys appear with the respective energies of $E_{J}=110\;\rm{meV}$, $E_{K}=220\;\rm{meV}$ and $E_{L}=260\;\rm{meV}$ above the global CBM of $I$.  According to previous calculations on electronic properties and thermoelectric performance of $\alpha$-Sb monolayer based on the constant relaxation-time approximation (CRTA)\cite{Pengbo2018,ywu2019} , which is a commonly employed approximation in the TE community, the $zT$ value of $\alpha$-Sb reaches 0.90 at room temperature (RT), which is higher than SnSe at RT and can be further optimized to 1.2 with tensile strain\cite{Wang2015d,ywu2019}. For $\beta$-Sb monolayer, in spite of its large band gap (1.25 eV) compared with conventional high-$zT$ materials, the CRTA-based calculations give a $zT$ value reaching 2.15 at RT, which is far superior to conventional thermoelectric materials\cite{Chen2017}. However, special care should be taken for the system with band convergence of multi-valley conduction bands or multi-peak valence bands, when evoking the CRTA, which is generally obtained from the deformation potential approximation (DPA) theory for non-polar semiconductors, by considering only the coupling between electrons and longitudinal acoustic (LA) phonons in the long-wavelength limit\cite{Xi2012,Qiao2014,Du2015,Witkoske2019}. 

In fact, the DPA theory might fail and misestimate the electron relaxation-time $\tau$ and then the intrinsic carrier mobility $\mu$ for non-polar semiconductors with band convergence for three reasons: \textcolor{black}{(1) Within the intravalley scatterings, the contribution to $\tau$ from phonon modes other than LA phonon is neglected\cite{Li2013}; (2) For band-convergence systems, the intervalley scattering of phonons may be considerably large\cite{Li2013,Liu2017,Sohier2019}; (3) The widely used deformation potentials in DPA method are calculated along specific directions which is approximately correct for simple parabolic valley/peak, but fails for highly anisotropic valleys/peaks\cite{Lang2016}.}
As an example, for monolayer stanene, the intrinsic carrier mobility $\mu$ predicted by the DPA theory is equal to $\sim10^6\;\rm{cm^{2}V^{-1}s^{-1}}$ at RT, but reduces to $2\sim3\times10^3\;\rm{cm^{2}V^{-1}s^{-1}}$ when considering full electron-phonon (\textit{el-ph}) couplings\cite{Nakamura2017}. 

To accurately calculate the electronic conductivity $\sigma$, which also determines the electronic thermal conductivity $\kappa_{e}$ according to the Wiedemann-Franz law, i.e. $\kappa_e=L\sigma T$,  here $L$ is the Lorenz number, the full \textit{el-ph} interactions in semiconductors should be investigated in details. The calculated phonon dispersions for $\beta$-Sb and $\alpha$-Sb are shown in Fig. S2 (c, d), and  the maximum phonon energies for $\beta$-Sb and $\alpha$-Sb are $20.90\;\rm{meV}$ and $20.92\;\rm{meV}$ respectively. Both monolayers possess phonon bandgaps around 13.0 meV with the respective values of 7.96 meV and 3.27 meV.  

\begin{figure}
\centering
\includegraphics[width=1\linewidth]{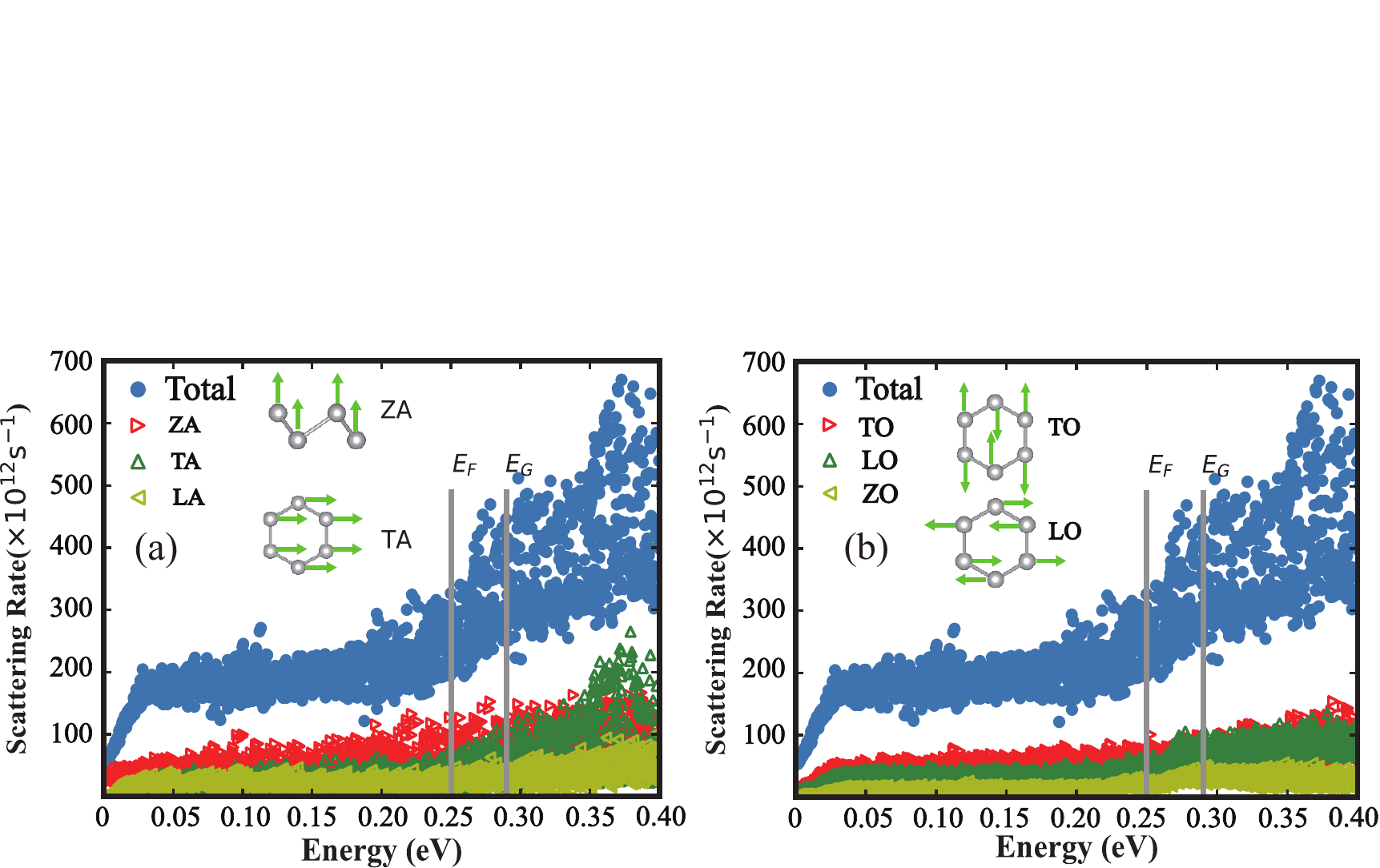}
\caption{The scattering rate of electrons in the $H$ valley with energies within $\sim0.4\;\rm{eV}$ of the CBM. (a) The contributions of acoustic branch phonons to the total scattering rate. \textcolor{black}{$E_F$ and $E_G$ are the position of $F$ and $G$ valleys above the CBM with $E_F=0.25$ eV and $E_G=0.29$ eV.} (b) The contributions of optical branch phonons to the total scattering rate. Atomic displacements of phonon modes are also shown in the figure. The sequences are according to the phonon energies seen in Fig. S2(c, d)}
\label{elsca}
\end{figure}

For convenience, $\beta$-Sb is investigated firstly. The mode-resolved \textit{el-ph} scattering rate ($1/\tau_{n\textbf{k}}$) for $\beta$-Sb near $H$ valley at 300 K within $0.40\;\rm{eV}$ above the CBM is shown in Fig.~\ref{elsca}.
For conduction electrons of ($n\textbf{k}$) in H valley, only the intravalley scatterings and the intervalley scatterings between degenerate valleys are possible, since the energy difference between CBM and F valley of $E_F$ is smaller than the maximum energy of phonons. The total \textit{el-ph} scattering rates denoted by blue dots in Figs.~\ref{elsca}(a,b) are approximately equal to $\sim2\times10^{14}\;\rm{s}^{-1}$, which is two orders larger than those of stanene\cite{Nakamura2017}. \textcolor{black}{By separating the phonon momenta near $\Gamma$ point and those away from $\Gamma$ point, the respective contributions from intravalley and intervalley scattering to the total scattering rates are listed in Table~\ref{intra-inter}, which reveals that the intervalley scattering overwhelms the intravalley scattering by at least one order of magnitude.} The large intervalley scattering was also reported for silicene and stanene\textcolor{black}{\cite{Nakamura2017}}, but it is more obvious in $\beta$-Sb, which is resulted from the six-fold degeneracy of CBM in the Brillouin zone for $\beta$-Sb as shown in Fig. S3(a). By comparison, we found that the contribution from optical phonons is comparable to that from acoustic phonons for $H$-valley electrons below $E_F$. Among them, the out-of-plane flexual ZA phonons around $\Gamma$ point contribute dominantly to the total \textit{el-ph} scatterings, which is similar to cases in group-IV buckled materials (silicene, germanene, stanene).  The buckled structure in antimonene monolayers increases the overlap of the electronic $p_z$ orbitals, thus electrons are more sensitive to the ZA phonons. In addition to the dominant contribution from flexual ZA phonons, LA, TO and LO phonons also contribute significantly to total \textit{el-ph} scatterings. 

As we know, the $\sigma_h$-symmetry in graphene ($D_{6h}$) and $\rm{MoS}_2$ ($D_{3h}$) restricts the lattice potential associated with the flexural displacement to be odd with respect to the in-plane mirror, leading to the vanishing of odd ZA phonons in the \textit{el-ph} matrix element according to the Mermin-Wagner theorem\cite{Fischetti2016,Nakamura2017}. Hence, the coupling of electrons and ZA phonons is suppressed and the DPA method based on the electron-LA phonon coupling works for these non-polar materials. However, the Mermin-Wagner theorem is not suitable for antimonene, since it is based on the assumption that the electron is near the high symmetry K point\cite{Wang2017}.

Based on the group-theory analysis, the selection rules for electron-phonon scattering can be derived by the direct product of the irreducible representation (irreps) of electronic initial states and the involved phonon modes\cite{Malard2009}, as follows, 
\begin{equation}
\Gamma^{f} \otimes \Gamma^{ph} \otimes \Gamma^{i}	\label{sim_selection_rule} = non\quad null\quad (null)
\end{equation}
where $\Gamma^{i,f,ph}$ represent the irreps for initial/final electron states and phonon modes respectively. The electron-phonon scattering is allowed/forbidden if the direct product gives a \textit{non null}/\textit{null} value. As Table S1 shows, the irreps for initial and final states near CBM of $\beta$-Sb are both A', belonging to C$_2$ group. Since the direct product, A'$\otimes$A'$\otimes$A'=non null, hence the selection rules only allow ZA and LA phonon modes involved in the intravalley scattering processes for H-valley electrons in $\beta$-Sb. For the intervalley electron-phonon scatterings between degenerate CBM in the Brillouin Zone as shown in Table S2, similar to the case of intravalley scatterings, only LA, ZA phonon modes are allowed to be involved restricted by the selection rules. Similar analysis can be applied for electron- and hole-phonon scatterings for monolayer $\beta$-Sb and stanene, as shown in Table S1 and Table S2, where the resulted selection rule for stanene is in good agreement with previous work\cite{Nakamura2017}. It should be noted that, according to Table.~\ref{intra-inter}, the intravalley scattering by ZA phonon mode is suppressed, although the ZA mode is allowed according to the selection rule, which is due to the failure to satisfy energy conservation for ZA modes with narrow range of phonon frequencies.

 

\begin{table}
\centering
\caption{intravalley scattering and intervalley scattering for electrons in $\beta$-Sb at CBM point and T = 300 K.}
\begin{tabular}{ccccc}
\hline
 Phonon mode & intravalley ($s^{-1}$) & intervalley($s^{-1}$) & Total($s^{-1}$)\\
\hline
 ZA & $3.72\times10^{-16}$ & $2.16\times10^{13}$ & $2.16\times10^{13}$ \\
 TA & $2.30\times10^{8}$ & $1.08\times10^{11}$ & $1.08\times10^{11}$  \\
 LA & $1.06\times10^{12}$ & $3.66\times10^{12}$ & $4.72\times10^{12}$ \\
 TO & $5.34\times10^{11}$ & $1.47\times10^{13}$ & $1.52\times10^{13}$ \\
 LO & $9.81\times10^{11}$ & $1.15\times10^{13}$ & $1.24\times10^{13}$ \\
 ZO & $9.69\times10^{11}$ & $3.96\times10^{12}$ & $4.93\times10^{12}$ \\
\hline
\end{tabular}
\label{intra-inter}
\end{table}



The intervalley scatterings of electrons from $H$ to $F$ valley via phonons with large momentums are possible when the H-valley electron energy is higher than $E_F$, and $H-G$ intervalley scatterings are allowed as well when the chemical potential is tuned to higher than $E_G$ by doping. As a result, the total \textit{el-ph} scattering rate $\tau$ increases abruptly near $E_F$ and increases further when $E>E_G$, as shown in Fig.~\ref{elsca}. The total \textit{el-ph} scattering rate around 0.38 eV is three times larger than that in the region $E<E_F$. By comparison, we found that, in the region where $E>E_F$ and $H-F$ intervalley scatterings are allowed, acoustic phonons contribute to the total \textit{el-ph} scatterings more than optical phonons do.  For pure $H-F$ intervalley scatterings of electrons with energies in the region $E_F$ and $E_G$, flexual ZA phonons dominate the total \textit{el-ph} scatterings, and the contributions from TA, TO and LO phonons increase when $E$ increases, and become comparable near $E_G$. When the chemical potential of $E$ increases larger than $E_G$ and further, the contribution from TA phonons increases more rapidly than ZA phonons, and becomes larger than ZA phonons near 3.7 eV. However, in this energy region where both $H-F$ and $H-G$ intervalley scatterings are allowed, the contributions from ZA, TO and LO phonons to total \textit{el-ph} $\tau$ are not negligible compared with the dominant TA phonons.

\begin{figure*}
\centering
\includegraphics[width=0.9\linewidth]{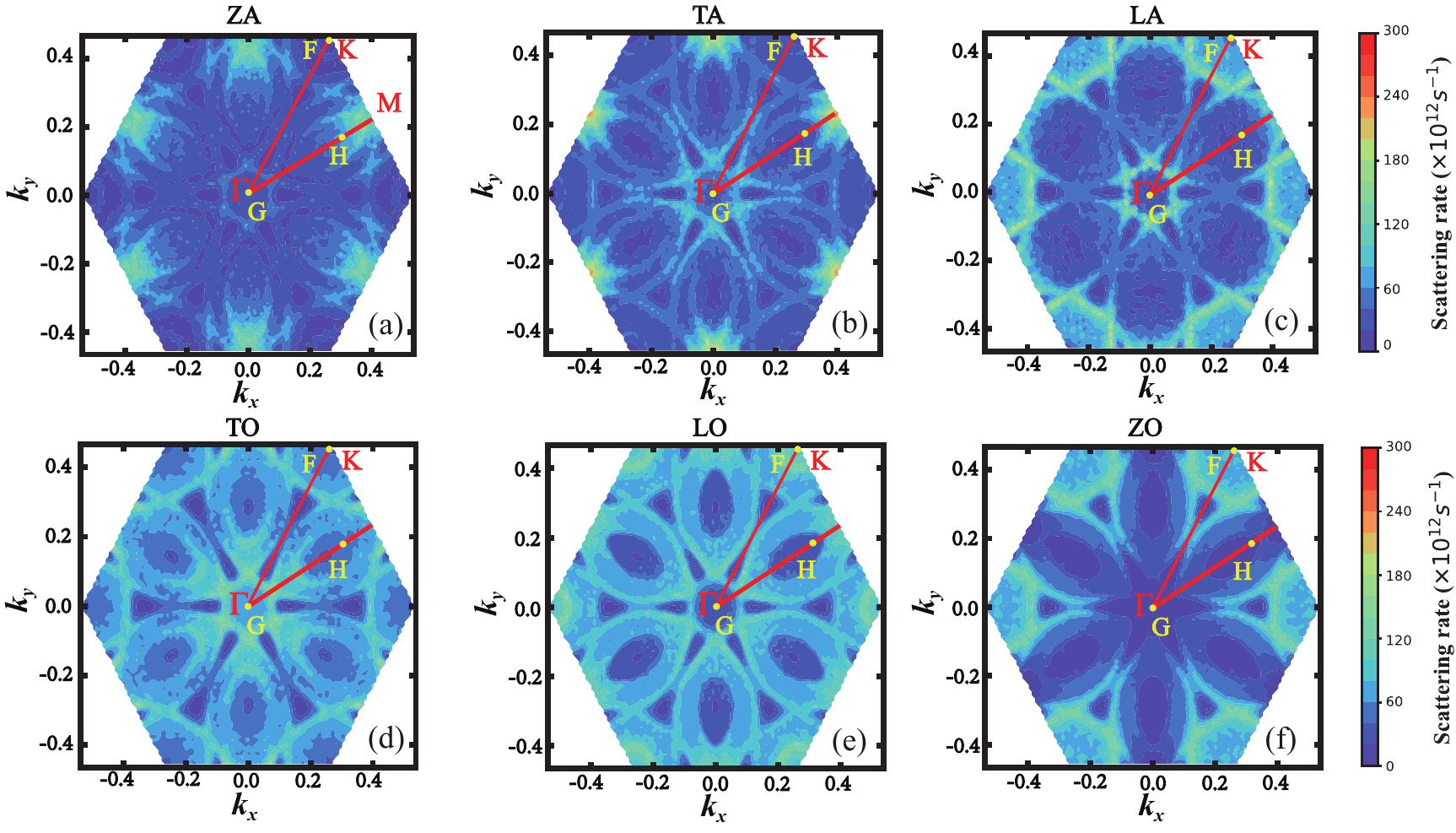}
\caption{The scattering rate of electrons in the conduction band for $\beta$-Sb as a function of wave vector $\textbf{k}$ at first Brillouin zone. The labels $\Gamma$, M and K represent the high symmetry point of wave vector $\textbf{k}$.}
\label{elscazone}
\end{figure*}

To further reveal the \textit{el-ph} scatterings of the conduction electrons via different phonon modes, the mode-resolved scattering rates of conduction electrons of $\beta$-Sb as a function of wave vector $\textbf{k}$ located \textcolor{black}{in the first Brillouin zone} are shown in Fig.~\ref{elscazone}. The distribution of the \textit{el-ph} couplings of conduction electrons for $\beta$-Sb also exhibits sixfold rotation-symmetry. For $H$-valley electrons, flexual ZA phonons dominate the intervalley scatterings. Differently, scatterings of $F$-valley electrons are dominated by LA, TO, LO and ZO phonon modes, and scatterings of $G$-valley electrons are dominated by LA and TO phonons. 

The mode-resolved scattering rates of holes for $\beta$-Sb are shown in Fig.~S4. Since the maximum of valence bands is well higher than the second peak, only intra-peak scatterings of holes are allowed for shallow doping.  As shown in Table S1, the irreps of initial and final electron states modes near VBM of $\beta$-Sb are both A, belonging to the small group C$_2$, and the irreps of ZA, TA and LA phonon modes are B, B and A respectively, belonging to the small group C$_2$ as well. In order to determine the dominant phonon modes involved in the intra-peak scatterings of valence holes, the analysis of the selection rules based on symmetry is performed. According to Eq.(\ref{sim_selection_rule}), the direct product of A$\otimes$A$\otimes$A is non null, suggesting the scattering between valence holes and LA phonon mode is allowed. However, the direct product of A$\otimes$B$\otimes$A is null, suggesting that \textit{el-ph} scatterings involved TA, ZA phonon modes are not allowed. Therefore, merely LA mode is involved in intra-peak \textit{el-ph} scattering near VBM for $\beta$-Sb, which is consistent with the numerical calculations of scattering rates shown in Fig. S4. And there is no inter-peak \textit{el-ph} interaction between degenerate VBM of $\beta$-Sb, since the VBM locates at $\Gamma$ point. Hence, in contrast to the general conclusion discovered in group-IV elemental materials (Silicene, Silicene, Stanene) in which the carriers (electrons and holes) in 2D semiconductors with $D_{3d}$ symmetry dominantly suffer from scatterings via ZA phonons, holes in $\beta$-Sb suffer dominantly from scattering via LA phonons.



\begin{figure*}
\centering
\includegraphics[width=0.8\linewidth]{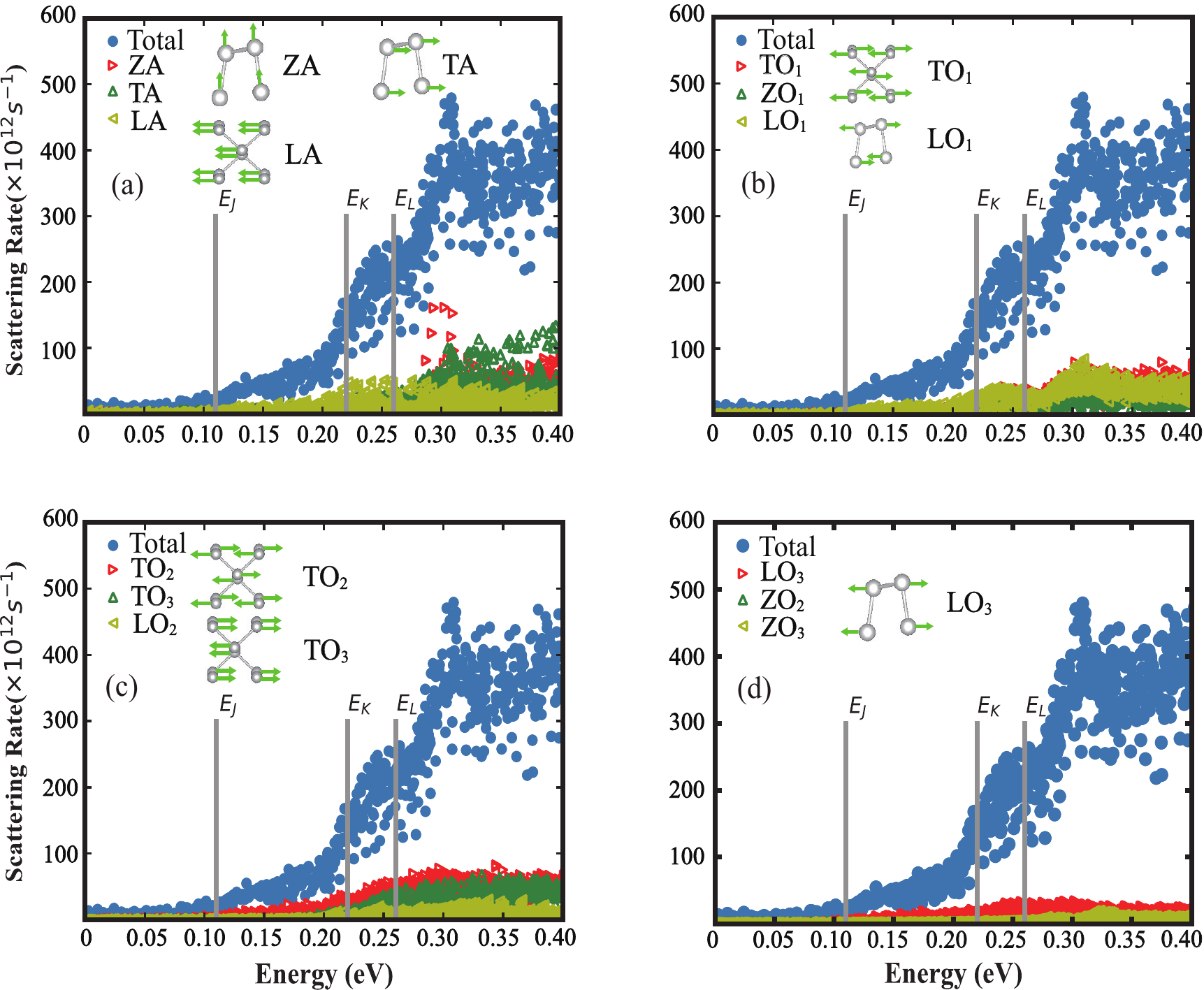}
\caption{The scattering rate of electrons in the $I$ valley for $\alpha$-Sb with energies within $\sim0.4\;\rm{eV}$ of the CBM. \textcolor{black}{$E_J$, $E_K$and $E_L$ are the position of $J$, $K$ and $L$ valleys above the CBM with $E_J=0.11$ eV, $E_K=0.22$ eV and $E_L=0.26$ eV.} (a) The contributions of acoustic branch phonons to the total scattering rate. (b-d) The contributions of optical branch phonons to the total scattering rate.}
\label{elscaalph}
\end{figure*}

Similarly, for $\alpha$-Sb, the mode-resolved \textit{el-ph} scattering rates ($1/\tau_{n\textbf{k}}$) for $I$-valley electrons at 300 K within $0.40\;\rm{eV}$ above the global CBM are calculated and shown in Figure.~\ref{elscaalph}. Compared with $\beta$-Sb, the carrier scattering rates for $\alpha$-Sb are much smaller within the energy range of $\sim0.1\;\rm{eV}$, which means that, the scatterings of $I$-valley electrons are weak. The mode-resolved scattering rates separated by intervalley and intravalley scatterings are also shown in Table~\ref{intra-inter-alpha}. For acoustic phonon branches, intervalley scattering by ZA phonons and intravalley scattering by LA phonons play important roles, similar to $\beta$-Sb. Besides, optical phonon modes (LO$_1$, TO$_2$, LO$_2$, LO$_3$, ZO$_2$ and ZO$_3$) also contribute significantly to the total scattering rates. When the chemical potential exceeds $E_J$, $I-J$ intervalley scatterings are allowed, and the total \textit{el-ph} scattering rates increase due to the enhanced intervalley scatterings by the combination of TO$_2$, TO$_3$ and LO$_3$ phonons. When the chemical potential is higher than $E_K$, $I-K$ intervalley scatterings are allowed, leading to the rapid increase of the total \textit{el-ph} scattering rates, mainly attributed to ZA, LA, TO$_1$, TO$_3$, and LO$_3$ phonons. When the chemical potential increases further and larger than $E_L$, an enhancement of the total \textit{el-ph} scattering rates with five times larger than those where only $I-J$ intervalley scatterings are allowed, can be identified. The considerable enhancement is mainly due to the additive scatterings via TA and ZA phonons, as shown in Fig.~\ref{elscaalph}(a). As shown in Table. S3, for an example, for the initial state of electron at (0, 0.323, 0) point, the $I-J$, $I-K$ and $I-L$ intervalley scatterings contribute $4.4\%$, $29.9\%$ and $9.7\%$, respectively, to the total \textit{el-ph} scattering rates, .

\begin{table*}
\scriptsize
\centering
\caption{intravalley scattering and intervalley scattering for electrons in $\alpha$-Sb at CBM point and T = 300 K.}
\begin{tabular}{cccccccc}
\hline
 Phonon mode & intravalley ($s^{-1}$) & intervalley($s^{-1}$) & Total($s^{-1}$) & Phonon mode & intravalley ($s^{-1}$) & intervalley($s^{-1}$) & Total($s^{-1}$)\\
\hline
 ZA & $5.56\times10^{-73}$ & $1.25\times10^{12}$ & $1.25\times10^{12}$ & TO$_2$ & $5.17\times10^{10}$ & $4.95\times10^{10}$ & $1.01\times10^{11}$\\
 TA & $4.46\times10^{8}$ & $2.45\times10^{8}$ & $6.91\times10^{8}$ & TO$_3$ & $3.84\times10^{5}$ & $1.62\times10^{7}$ & $1.65\times10^{7}$\\
 LA & $6.37\times10^{11}$ & $2.60\times10^{7}$ & $6.37\times10^{11}$ & LO$_2$ & $2.99\times10^{11}$ & $2.34\times10^{8}$ & $2.99\times10^{11}$\\
 TO$_1$ & $7.66\times10^{6}$ & $8.60\times10^{9}$ & $8.60\times10^{9}$ & LO$_3$ & $8.88\times10^{6}$ & $1.13\times10^{11}$ & $1.13\times10^{11}$\\
 ZO$_1$ & $1.46\times10^{9}$ & $1.72\times10^{8}$ & $1.63\times10^{9}$ & ZO$_2$ & $1.81\times10^{11}$ & $6.46\times10^{7}$ & $1.81\times10^{11}$\\
 LO$_1$ & $2.68\times10^{11}$ & $7.92\times10^{11}$ & $1.06\times10^{12}$ & ZO$_3$ & $6.74\times10^{6}$ & $1.73\times10^{11}$ & $1.73\times10^{11}$\\
\hline
\end{tabular}
\label{intra-inter-alpha}
\end{table*}

The mode-resolved \textit{el-ph} scattering rates of holes along $\Gamma-Y$ direction within $\sim0.4$ eV below the VBM for $\alpha$-Sb are shown in Figure~S7. The energy level of VBM for $\alpha$-Sb is well above the second peak as well, only the intravalley scatterings and the intervalley scatterings between degenerate valleys are allowed. The total \textit{el-ph} scattering rates of holes are mainly attributed to the combination of ZA, LA, LO$_1$ and TO$_2$ phonons.

\textcolor{black}{Based on the calculated mode-resolved \textit{el-ph} scattering rates of electrons and holes, their carrier mobility can be calculated by solving the Boltzmann transport theory according to,}

\begin{equation}
\mu_{\alpha \beta}=\frac{-e}{n_{e(h)}\Omega}\sum_{n\in CB(VB)}\int\frac{d\textbf{k}}{\Omega_{BZ}}  \frac{\partial{f^{0}_{n\textbf{k}}}}{\partial{\epsilon_{n\textbf{k}}}}v_{n\textbf{k},\alpha}v_{n\textbf{k},\beta}\tau_{n\textbf{k}}
\label{mu_ab}
\end{equation}

where $n_{e(h)}$ is the electron(hole) density, $\Omega$ and $\Omega_{BZ}$ denote the volume of the unit cell and the first Brillouin zone, respectively, $v_{n\textbf{k}, \alpha}=h^{-1}\partial{\epsilon_{n\textbf{k}}}/\partial{k_{\alpha}}$ is the velocity of the single-particle $n\textbf{k}$ electron along $\alpha$ direction. Fine $\textbf{k}$ and $\textbf{q}$ meshes are needed for converging the k-integral in Eq.~(\ref{mu_ab}) and the q-integral for $\tau$. In order to investigate the temperature-dependent mobilities, we first carried out the \textit{ab initio} molecular dynamics (AIMD) calculations to simulate the behavior of monolayer $\beta$-Sb and $\alpha$-Sb at 300 K, 500 K and 600 K respectively, and the results are shown in Fig.~S4, which confirms the thermal stabilities of both $\alpha$- and $\beta$-Sb at these temperatures.

\begin{figure*}
\centering
\includegraphics[width=1\linewidth]{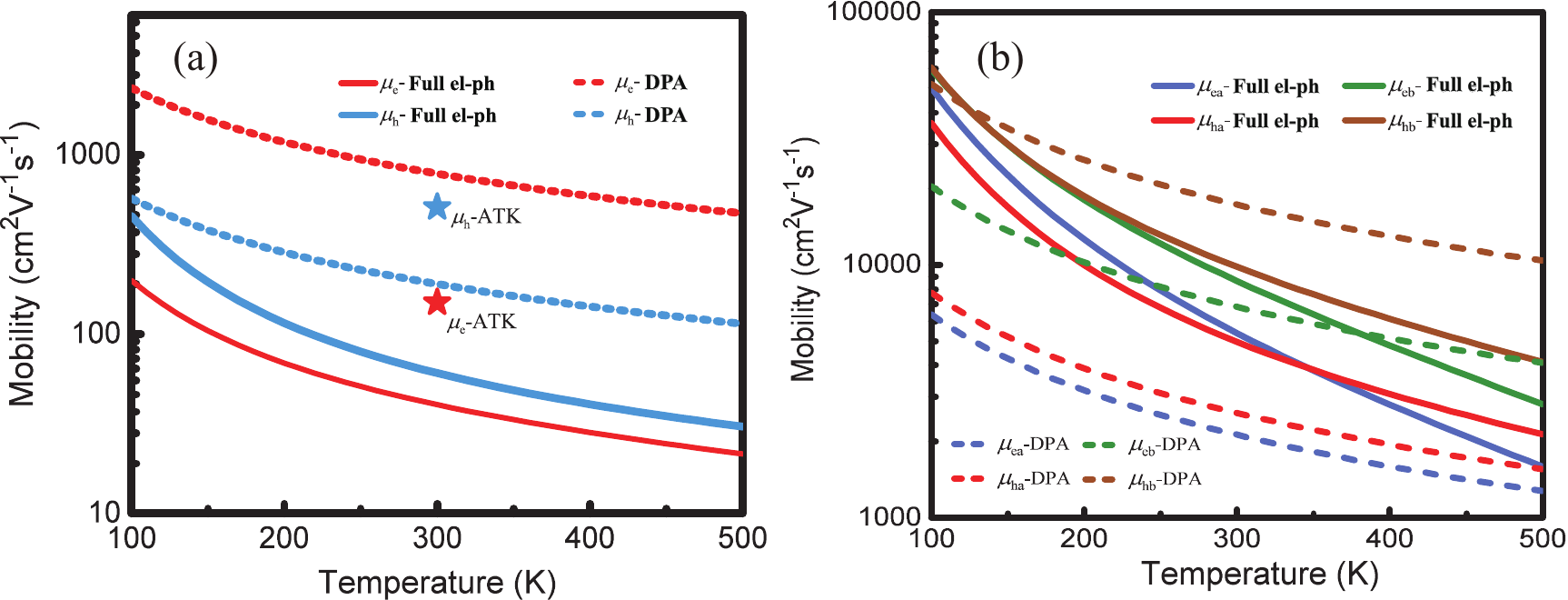}
\caption{\textcolor{black}{The calculated carrier mobility for electrons ($\mu_e$) and holes ($\mu_h$) for (a) $\beta$-Sb and (b) $\alpha$-Sb.The dashed line and solid line represent the results under DPA and full \textit{el-ph} coupling method respectively. The pentacle represents the results calculated using ATK package considering intravalley scatterings by three acoustic branches\cite{Wang2017}.} }
\label{mob}
\end{figure*}

The calculated intrinsic temperature-dependent carrier mobilities for $\beta$-Sb and $\alpha$-Sb ranging from 100 K to 500 K are shown in Figure.~\ref{mob} as solid lines. The carrier mobilities decrease gradually as the increase of the temperature, which is due to more phonons excited at higher temperature. At 300 K, the calculated electron and hole mobilities for $\beta$-Sb are $40\;\rm{cm^{2}V^{-1}s^{-1}}$ and $61\;\rm{cm^{2}V^{-1}s^{-1}}$, respectively, when considering full \textit{el-ph} interactions for electrons and holes. As a comparison, the calculated electron and hole mobilities at 300 K given by the DPA method which only considers the coupling between electrons/holes and LA phonons, are $785\;\rm{cm^{2}V^{-1}s^{-1}}$ and $205\;\rm{cm^{2}V^{-1}s^{-1}}$, respectively, which are 19.6 and 3.4 times larger than those calculated based on full \textit{el-ph} interactions. The reason why the difference ratio for holes is much smaller is because the intravalley scatterings of holes via LA and TO phonons play the major role in determining hole mobility for $\beta$-Sb, which is similar to the underlying mechanism of the DPA method.
When only considering the interactions between long-wavelength acoustic branches of phonons and the electrons/holes near the CBM and VBM, and excluding the intervalley-scattering events using the Atomistix ToolKit (ATK) package, the calculated electron and hole mobilities at 300 K are $150\;\rm{cm^{2}V^{-1}s^{-1}}$ and $510\;\rm{cm^{2}V^{-1}s^{-1}}$, respectively\cite{Wang2017}, which are 4 and 8 times larger than this work. 
The deviation reveals the importance of intervalley scattering to the carrier mobilities in $\beta$-Sb. 

The calculated electron and hole carrier mobilities for $\alpha$-Sb at 300 K along $a$ and $b$ directions (as defined in Fig. S1) considering full \textit{el-ph} interactions are $5370\;\rm{cm^{2}V^{-1}s^{-1}}$ and $8635\;\rm{cm^{2}V^{-1}s^{-1}}$ , and $4961\;\rm{cm^{2}V^{-1}s^{-1}}$ and $9818\;\rm{cm^{2}V^{-1}s^{-1}}$, respectively. As a comparison, the modified DPA method gives electron and hole mobilities at 300 K along $a$ and $b$ directions of $2125.62\;\rm{cm^{2}V^{-1}s^{-1}}$ and $6818.51\;\rm{cm^{2}V^{-1}s^{-1}}$, and $2593.66\;\rm{cm^{2}V^{-1}s^{-1}}$ and $17313.09\;\rm{cm^{2}V^{-1}s^{-1}}$. The neglect of scatterings from optical phonons also results in the overestimate in carrier mobilities. It should be noted that the calculated mobilities using the DPA method are smaller than those considering full \textit{el-ph} scatterings especially for the electron mobilities along $a$ direction, as shown in Fig.~\ref{mob}(b). The underlying mechanism can be understood roughly by the fact that, in DPA method, all the transport electrons suffer from scatterings via LA phonons, which are dominant in intravalley scatterings and strong as shown in Table~\ref{intra-inter-alpha}, however in antimonene a large part of transport electrons suffer from scatterings via TA or other phonon modes with weak scattering strengths, leading to the underestimate of mobilities compared with those considering full \textit{el-ph} scatterings. 

\begin{figure*}[ht!]
\centering
\includegraphics[width=0.8\linewidth]{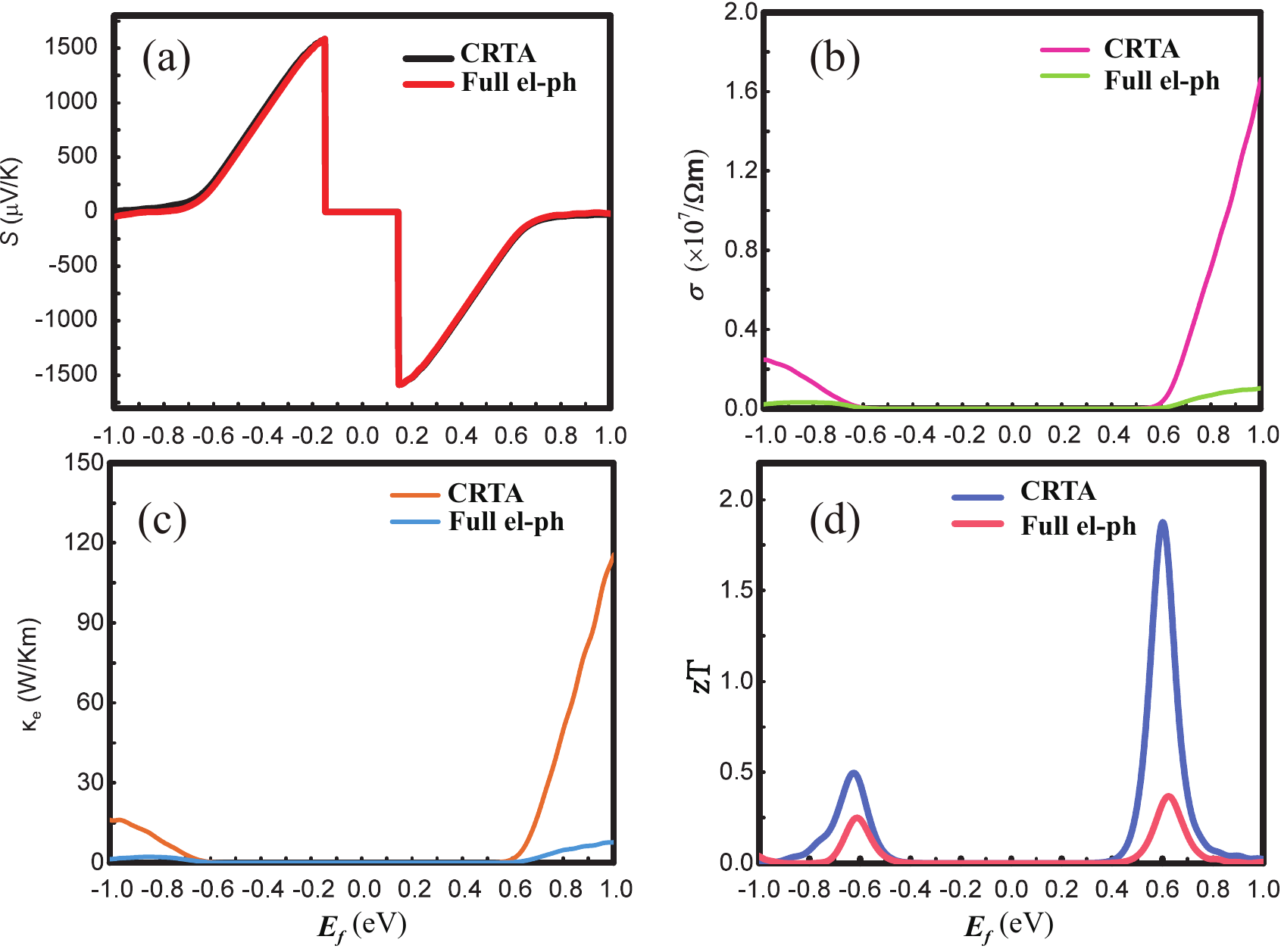}
\caption{The (a) seebeck coefficients (b) electrical conductivity (c) electronic thermal conductivity and (d) $zT$ value as a function of chemical potential at 300 K for $\beta$-Sb.}
\label{betathermoelectric}
\end{figure*}

\begin{figure*}[ht!]
\centering
\includegraphics[width=0.8\linewidth]{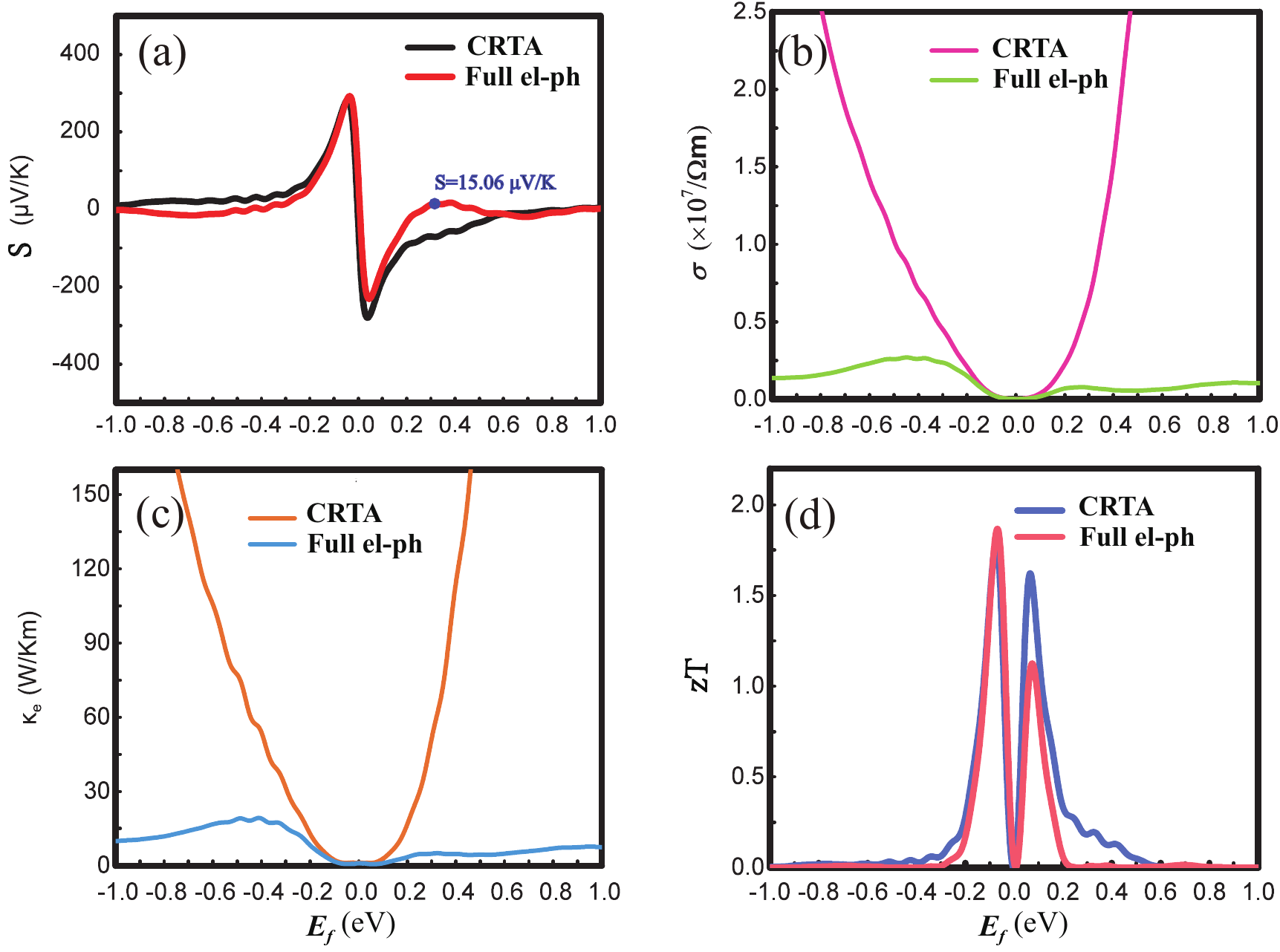}
\caption{The (a) seebeck coefficients (b) electrical conductivity (c) electronic thermal conductivity and (d) $zT$ value as a function of chemical potential at 300 K for $\alpha$-Sb along $a$ direction.}
\label{alphathermoelectric}
\end{figure*}

The Seebeck coefficients $S$, electronic conductances $\sigma$ and electronic thermal conductivities $\kappa_e$ based on the rigid-band approximations for $\beta$- and $\alpha$-Sb are calculated and shown in Figs.~\ref{betathermoelectric}(a-c) and Figs.~\ref{alphathermoelectric}(a-c), respectively. For comparison, $S$, $\sigma$ and $\kappa_e$ are calculated based on the constant relaxation time approximation (CRTA) and full \textit{el-ph} interactions respectively, and both CRTA and full \textit{el-ph} calculations are performed using the BoltzTrap2 package\cite{Madsen2018}. For CRTA calculations, the constant electronic relaxation time (new-$\tau$) is obtained according to the DPA method as listed in Table S4. For $\beta-Sb$, the Seebeck coefficient $S$ is insensitive to electron-phonon couplings and reaches nearly $1500\;\rm{\mu V/K}$ when the chemical potential is $\pm 0.13\;\rm{eV}$. The impact of the full \textit{el-ph} interactions on $S$ is mediated through the energy-dependent carrier lifetime $\tau(E)$, which is described by the Mott relation\textcolor{red}{\cite{Sun2015,Wei2015,Liang2016,Yang2016}},

\begin{equation}\label{Mott}
S=-\frac{\pi^{2}}{3} \frac{k_{B}^{2} T}{e}\left[\frac{\partial \ln N(E)}{\partial E}+\frac{\partial \ln \tau(E)}{\partial E}\right]_{E_{f}}
\end{equation}

Where $N(E)$ and $\tau(E)$ are the energy dependent DOS and electronic relaxation time. The Mott relation shown as Eq.~(\ref{Mott}) suggests that, a weakly-energy-dependent relaxation time $\tau$ influences little on the $S$. As shown in Fig.~\ref{elsca} and Fig.S(4), $1/\tau$  maintains basically unchanged with the energy shift, leading to the relative insensitivity to \textit{el-ph} scatterings. However, the electronic conductance of $\sigma$ strongly depends on the \textit{el-ph} couplings via $\sigma=ne\mu$, where $n$ is the carrier concentration.



 
The total thermal conductivities are the sum of the electronic thermal conductivities $\kappa_e$ and the lattice thermal conductivities $\kappa_l$, i.e. $\kappa=\kappa_e+\kappa_l$. At low temperature, phonons are dominantly scattered by impurities and boundaries, and $\kappa_l$ is proportional to the square of the temperature, $\kappa_l\propto T^2$ in two dimensions according to Debye model. When the temperature is well larger than Debye temperature ($\Theta_D$), Umklapp processes of phonon-phonon scattering dominates and $\kappa_l\propto 1/T$\cite{Kim2016}. The calculated $\Theta_D$ for $\beta$-Sb and $\alpha$-Sb are 98.4 K and 230 K (272 K) respectively\cite{Guo2017,Pengbo2018}. Thus at RT, the lattice thermal conductivity $\kappa_l$ inversely depends on temperatures. In addition, the lifetimes of phonons by electron scatterings as shown in Fig.~S8, are considerably small and can be neglected according to the Mattiessen's Rule. Hence, at RT the total $\kappa$ can be roughly estimated by the lattice $\kappa_l$ considering only three-phonon scatterings, which has been calculated by solving the semiclassical Boltzmann transport equation (BTE) in our previous works\cite{Pengbo2017,Pengbo2018}. Here we use the values of $\kappa_l$ at RT therein.

The dimensionless figure of merit $zT$ for $\beta$- and $\alpha$-Sb are subsequently calculated and shown in Fig.~\ref{betathermoelectric}(d) and Fig.~\ref{alphathermoelectric}(d), respectively. At 300 K, a maximum value of $zT\sim1.88$ at $E_f=0.62 \;\rm{eV}$ for $\beta$-Sb under CRTA is realized, compared with conventional thermoelectric materials like  Bi$_2$Te$_3$ ($1.2$)\cite{Poudel2008}, PbTe ($0.30$) \cite{Qinyong2013}, SnSe ($0.70$) \cite{Wang2015d}. However, when considering full \textit{el-ph} interactions, the maximum $zT$ value decreases by 5.1 times  due to the sharp decrease of electronic conductivity $\sigma$.

For $\alpha$-Sb along $a$ direction (Fig.~\ref{alphathermoelectric}), the Seebeck coefficient S is also insensitive to the energy-dependent \textit{el-ph} coupling near $E_f=0$ as $\beta$-Sb. However, when the chemical potential is larger than 0.2 eV, i.e. $E_f>0.2\;\rm{eV}$, the full \textit{el-ph} coupling introduces strongly-energy-dependent relaxation time, as shown in Fig. \ref{elscaalph}, thus the $S$ curve with full \textit{el-ph} scattering shifts from the CRTA-$S$ curve at $E_f>0.2\;\rm{eV}$ for n-type system, according to the Mott relation. $S$ reaches maximum of $15.06\;\mu \mathrm{V} / \mathrm{K}$ at $E_f=0.316\;\rm{eV}$.
As shown in Fig.~\ref{alphathermoelectric}(c,d), the curves of $\sigma$ and $\kappa_e$ with CRTA near $E_f=0$ are in good agreement with the full \textit{el-ph} coupling results, which is due to the constant relaxation time obtained using modified DPA method consistent with the full electron-phonon coupling theory at CBM (VBM)($\sim 10^{-13}\;\rm{s}$). However, with the increase of $E_f$, $\sigma$ and $\kappa_e$ with CRTA deviate far from the those calculated by the full electron-phonon couplings, resulted from the significant underestimation of scattering rates. 

Fig.~\ref{alphathermoelectric}(d) shows the calculated $zT$ for $\alpha$-Sb using these two methods. The CRTA method gives the maximum $zT$=1.78 at $E_f=-0.08\;\rm{eV}$, which is very close to that calculated by the full \textit{el-ph} coupling, but higher than that obtained by using the traditional DPA method\cite{Pengbo2018,ywu2019}. It is attributed to the fact that the traditional DPA method gives rough carrier mobilities for anisotropic materials, since it assumes that moving carriers are only scattered by phonons along the same direction. Hence, the calculated relaxation time is smaller by an order for holes along $a$ direction (Table. S4), leading to the underestimate of $\sigma$ near $E_f=0$. For $E_f>0$, the maximum $zT$ reaches 1.62 in the CRTA theory, which is larger than $zT=1.13$ based on the full \textit{el-ph} coupling method, mainly due to the neglect of the intervalley scattering effects. The case for $b$ direction is shown in Fig.S9. Thus, the CRTA method is a good approximation to describe the carrier transport for $\alpha$-Sb with shallow doping, as well as TE performance of p-type $\alpha$-Sb.

In summary, we  systematically analyze the electron-phonon interactions in $\beta$- and $\alpha$-Sb. For electrons in $\beta$-Sb, the ZA phonon scattering is dominant similar to the 2D stanene and the intervalley channels play an important role to the scattering of electrons. The calculated  electron mobilities are $40\;\rm{cm^{2}V^{-1}s^{-1}}$ at room temperature, which are reasonably smaller than the DPA results. For holes in $\beta$-Sb, DPA method gives a good approximation due to the fact that the LA phonons dominate the scattering process. As a result, the maximum $zT$ value at room temperature experiences almost 5.1 times reduction to 0.37 when injecting electrons. For $\alpha$-Sb, the carriers at band edge experience little scattering by phonons and LA branch contributes most. Thus, $\alpha$-Sb maintains ultrahigh carrier mobilities in a wide temperature range and the DPA method provides a good approximation. However, including the intervalley scattering effect decreases the maximum $zT$ value to 1.13 for n-type $\alpha$-Sb along $a$ direction. Therefore, in the multi-band systems, the $zT$ value can be overestimated for TE application using the constant relaxation time from DPA and the intervalley scattering needs careful considerations .

\section*{Acknowledgements}
This work is supported by the National Natural Science Foundation of China under Grants No. 11374063 and 11674062, the National Key R$\&$D Program of China (2017YFA0303403), the Shanghai Municipal Natural Science Foundation under Grant No. 19ZR1402900, the Natural Science Foundation of Jiangsu Province under grant No. BK20180456 and Fudan University-CIOMP Joint Fund (FC2019-006).

\section*{Conflict of Interest}
The authors declare no conflict of interest.

\section*{Keywords}
Thermoelectrics, Band convergence, Electron-phonon scattering, Carrier mobility, zT value


%

\end{document}